\documentclass[preprint,showpacs,preprintnumbers,amsmath,amssymb,prb]{revtex4}
\usepackage{epsfig,amsmath,amssymb,latexsym}
\newcommand{\bea}{\begin{eqnarray}}
\newcommand{\eea}{\end{eqnarray}}
\newcommand{\eg}{e.g.\! }

\begin{document}
\title{Analysis of optical properties of strained semiconductor quantum dots for electromagnetically  induced transparency}% Force line breaks with \\
%
%
%
%
% D. Barettin, J. Houmark, B. Lassen, M. Willatzen, T. R. Nielsen, J. M\o rk, and A.-P. Jauho
% Mads Clausen  Institute for Product Innovation,  University of Southern Denmark, 6400 S\o nderborg, Denmark
% Department of Micro- and Nanotechnology, Technical University of Denmark, 2800 Kgs Lyngby, Denmark
% Department of Photonics Engineering, Technical University of Denmark, 2800 Kgs. Lyngby, Denmark
%
\author{D. Barettin}
\email[Corresponding author: ]{daniele@mci.sdu.dk} \affiliation{Mads
Clausen  Institute for Product Innovation, University of Southern
Denmark, 6400 S\o nderborg, Denmark}
\author{J. Houmark}
\affiliation{DTU Nanotech -- Department of Micro- and
Nanotechnology, Technical University of Denmark, DK-2800 Kongens
Lyngby, Denmark}
\author{B. Lassen} \affiliation{Mads Clausen
Institute for Product Innovation, University of Southern Denmark,
6400 S\o nderborg, Denmark}
\author{M. Willatzen}
\affiliation{Mads Clausen  Institute for Product Innovation,
University of Southern Denmark, 6400 S\o nderborg,
Denmark}
\author{T. R. Nielsen}
\affiliation{DTU Fotonik, Department of Photonics Engineering,
Technical University of Denmark, Building 343, DK-2800 Kongens
Lyngby, Denmark}
\author{J. M\o rk}
\affiliation{DTU Fotonik, Department of Photonics Engineering,
Technical University of Denmark, Building 343, DK-2800 Kongens
Lyngby, Denmark}
\author{A.-P. Jauho} \affiliation{DTU Nanotech --
Department of Micro- and Nanotechnology, Technical University of
Denmark, DK-2800 Kongens Lyngby, Denmark} \affiliation{Department of
Applied Physics, Helsinki University of Technology, P. O. Box 1100,
FI-02015 HUT,
Finland}%

\date{\today}% It is always \today, today,
\begin{abstract}
Using multiband $\vec{k}\cdot\vec{p}$ theory we study the size and
geometry dependence on the slow light properties of conical
semiconductor quantum dots. We find the $V$-type scheme for
electromagnetically induced transparency (EIT) to be most favorable,
and identify an optimal height and size for efficient EIT operation.
In case of the ladder scheme, the existence of additional dipole
allowed intraband transitions along with an almost equidistant
energy level spacing adds additional decay pathways, which
significantly impairs the EIT effect. We further study the influence
of strain and band mixing comparing four different
$\vec{k}\cdot\vec{p} $ band structure models. In addition to the
separation of the heavy and light holes due to the biaxial strain
component, we observe a general reduction in the transition
strengths due to  energy crossings in the valence bands caused by
strain and band mixing effects. We furthermore find a non-trivial
quantum dot size dependence of the dipole moments directly related
to the biaxial strain component. Due to the separation of the heavy
and light holes the optical transition strengths between the lower
conduction and upper most valence-band states computed using
one-band model and eight-band model show general qualitative
agreement, with exceptions relevant for EIT operation.
\end{abstract}
%Use showkeys class option if keyword
\maketitle

\section{Introduction}
InAs/GaAs quantum dot structures have recently received much
attention due to their relevance for optoelectronic devices.
\cite{piprek} The discrete level nature of quantum dots offer important advantages over bulk and quantum well material for applications in conventional devices like lasers and optical amplifiers, as well as devices for all-optical signal processing. However, quantum dots also give the possibility of realizing solid-state implementations of effects and functionalities so far only demonstrated in atomic systems. For instance, the practical exploitation of slow light effects such as electromagnetically induced transparency (EIT) demonstrated in ultracold atoms~\cite{art:hau_nat} is naturally pursued using quantum dots. \cite{art:hasnain_optical_buffer} Important aspects
in this context are to accurately model the optical properties of
the quantum dots and to consider the influence of the lattice mismatch induced strain field and its effects on the bandstructure and eigenstates. Apart from Refs.~[\onlinecite{art:michael_apl, art:chow_jmo}] that focus on the many-body aspects of EIT operation most theoretical investigations of EIT induced slow light~\cite{art:Kim_slolight_condmat,art:Janes_2008,art:PKN_opt_express, art:PLH_APL_2009} have been based on simple quantum dot models that do not take into account important contributions from \eg biaxial strain.

%They have been proposed as components in devices for
%controlling the emission pattern of phased array
%antennas,\cite{art:Mork_laser_photon_2008}~as ultrafast optical
%amplifiers( citation needed) or in all--optical
%switches\cite{art:Wada_ultrafast_optics}~\eg as active media in
%buffers based on slow-light phenomena, utilizing either
%electromagnetically induced
%transparency\cite{art:hasnain_optical_buffer} (EIT) or population
%oscillation.\cite{art:chuang_optical_buffer_CPO} Important aspects
%in this context are to accurately model the optical properties of
%the quantum dots and to consider the influence of the lattice mismatch induced strain field
%and its effects on the bandstructure and eigenstates.

In this paper, we use $\vec{k}\cdot\vec{p}$ theory\cite{Kane,
Kittel} to determine the bandstructure of conical quantum dots. A
first popular $\vec{k}\cdot\vec{p}$ multiband calculation scheme for
bulk materials is due to Luttinger and Kohn
\cite{Luttinger1,Luttinger} which later on was extended to
heterostructures by an ad hoc symmetrization
procedure.\cite{Bastard88} In order to overcome this ad hoc
procedure, Burt formulated the so-called exact envelope function
method \cite{Burt1,Burt2} and soon after Foreman\cite{Foreman1} used
this method to derive a six-band model for the valence bands of
zincblende heterostuctures. Pokatilov et al.\cite{Pokatilov} have
provided an eight-band model for the conduction and the valence
bands. They studied quantum dots using a spherical
approximation and compared their model, based on exact
envelope-function theory, against the usual symmetrized approach.
The asymmetry parameter present in the Burt-Foreman formalism but not in the Luttinger-Kohn
formalism is shown to lead to changes of approximately $\pm 25$ meV in the electronic bandstructures
of InAs/GaAs. \cite{Pokatilov}

Optoelectronic properties of InGaAs zincblende quantum-dot with
varying shape and size based on $\vec{k}\cdot\vec{p}$ theory have
already been studied by Schliwa et al.\cite{Bimberg1} and Veprek et
al..\cite{Veprek} However, so far, only a small selection of all
possible transitions have been studied. In this work we apply the
eight-band model based on Burt-Foreman formalism derived in Ref.~[\onlinecite{Pokatilov}] to
zincblende InAs/GaAs conical quantum dots. We choose the conical geometry because (i) it introduces important symmetry lowering effects
not present in the earlier spherical models, and (ii) it mimicks closely the structures realized
in experimental systems, e.g., in Ref.~[\onlinecite{Inoue}]. We study the  size and
shape dependence. In particular, we show that some of the, for EIT yet
unexplored, interband transitions are highly relevant.
Furthermore, we investigate the effect of strain and band-mixing
between the conduction band and valence bands by comparing four
different $\vec{k}\cdot\vec{p}$ models. Although the previous $\vec{k}\cdot\vec{p}$  bandstructure
studies do include these effects, their impact on optical properties, such as EIT, have
not yet been investigated.

The paper is organized as follows. In Sec. II the theory for EIT,
bandstructure calculations and dipole moments is introduced. In Sec. III we
present results for the bandstructure and dipole moments calculations,
addressing the volume and shape effects, while in Sec. IV these results are
used to develop level schemes that lead to an efficient EIT operation. The paper is concluded in Sec. V.

\section{Theory}

\subsection{EIT}

EIT refers to an artificially created spectral region of transparency
in the middle of an absorption line due to the destructive
quantum interference arising from two transitions
in a three-level system.\cite{art:Harris_prl_1990, art:Harris_phys_today_1997}
By virtue of the Kramers-Kr\"{o}nig relations a dip in absorption is accompanied by a large positive slope of the refractive index which translates into a reduced group velocity in vicinity of the resonance.
Experimental realization of EIT in semiconductor structures has been achieved for the case of quantum wells Ref.~[\onlinecite{PhysRevLett.84.1019, art:philips_transient_EIT_prb_2004, Kang:08}], whereas for quantum dots (QDs), to the best of our knowledge, only two reports exist.\cite{art:Saulius_apl_2008,art:HAM_EIT_optexp2009}

\begin{figure}[ht]
    \begin{center}
    \includegraphics[width=9.5cm]{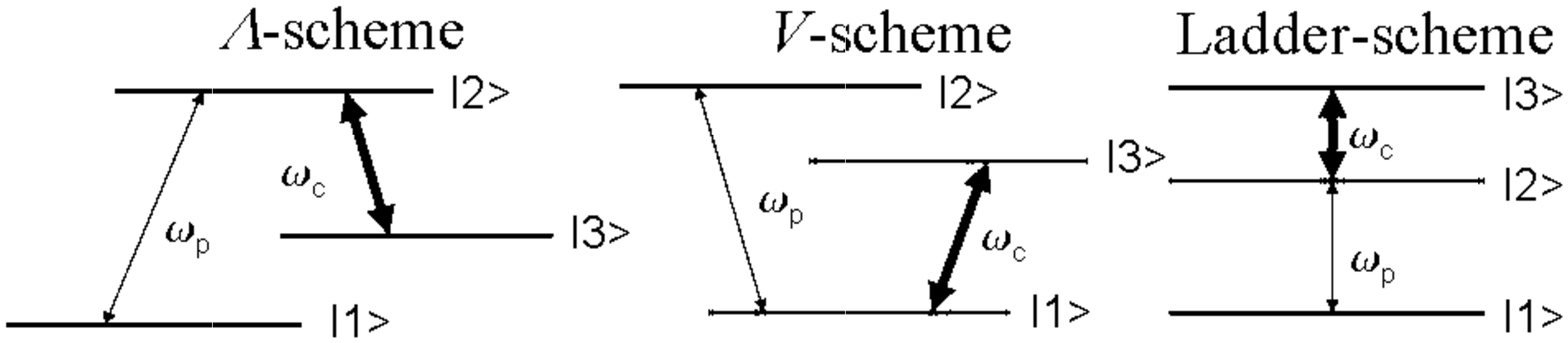}
    \end{center}
    \caption{The three generic EIT schemes. $\omega_p$ and $\omega_c$ refer to the frequency of the probe and coupling transitions, respectively.}
    \label{EIT_schemes}
    
 \end{figure}

EIT effects in quantum dots and wells have mostly been studied using models that are rooted in atomic physics assuming an archetypical EIT configuration in an ideal medium.\cite{art:hasnain_optical_buffer, art:Kim_slolight_condmat, art:Janes_2008, art:PKN_opt_express} These models truncate the number of active levels in the system to consider only the 3 levels being addressed by the laser fields. The generic EIT setup relies on a coupling and a probe laser driving separate transitions (see Fig.~\ref{EIT_schemes}). In case of the $\Lambda$-scheme, the ground state $|1\rangle$ has the
same parity as state $|3\rangle$, and the transition is thereby
dipole forbidden. State $|2\rangle$ is of the opposite parity and
dipole coupled to both $|1\rangle$ and $|3\rangle$. An intense near
resonant continuous wave electromagnetic field, termed the coupling field, drives
the $|2\rangle$-$|3\rangle$ transition. Another also near resonant,
but much weaker probe field, is applied to the
$|1\rangle$-$|2\rangle$ transition.

The archetypical EIT schemes are native to the atomic physics literature and generally one seeks to avoid excitation of carriers. In semiconductors such non-carrier exciting schemes involve a interband probe transition along with an intraband coupling transition. The frequency for the coupling field lies within the deep infrared, a range for which high intensity laser operation is very difficult. Carrier exciting schemes on the other hand involve two interband transitions, where the coupling field is effectively pumping carriers into the conduction band. The carriers interact via the Coulomb force, EIT schemes of this kind have been shown to be favorable when the effect of such many-body interactions are included.\cite{art:JHN_PRB_EIT,art:michael_apl,art:chow_jmo}

%Achieving the sufficient conditions for EIT. Most notably circumventing the effects of inhomogenous broadening (we have suggested a V-type scheme to overcome this obstacle) as well as high intensity laser operation at wavelengths around near infrared.

 The study of pulse propagation in a semiconductor slow light medium generally involves solving the coupled Maxwell-Bloch equations (see \eg Ref.~[\onlinecite{art:TRN_APL_2009}]). However, under certain circumstances an analysis of the steady state properties of the semiconductor Bloch equations (SBE) alone is adequate.~\cite{art:JHN_PRB_EIT} If this is the case then the linear optical response extracted via the SBE will be directly linked to the propagation characteristics of a wavepacket traveling in an optically thick system.

A relevant figure of merit for slow light operation is the slowdown factor $S$, which is defined as the ratio of the velocity of light in vacuum to the group velocity of a wavepacket traveling through the slow light medium:
\begin{eqnarray}
    \label{slowdown_factor} S(\omega) = \frac{c_0}{v_g} = n + \omega \frac{\partial n }{\partial \omega}\, ,
\end{eqnarray}
where $c_0$ is the speed of light in vacuum, and $n = {\rm Re}  \{        [\epsilon_b + \chi(\omega) ]^{\frac{1}{2}} \}$ is the refractive index, with $\epsilon_b$ being the background permittivity and $\chi(\omega)$ the susceptibility of the active material. The maximum slowdown is found at the frequency for which the slope of the refractive index is largest. Notice that the slowdown factor obtained away from resonance ($\frac{\partial n}{\partial \omega} \approx 0$) is given by the background refractive index.
The first order susceptibility, or linear system response, is found from the
induced macroscopic polarization $P(\omega)$ as $\chi(\omega)  =
\frac{P(\omega)}{\epsilon_0 E_\mathrm{p}(\omega)}$, where
$\epsilon_0$ is the vacuum permittivity and $E_\mathrm{p}(\omega)$
is the amplitude of the probe field. In turn the time resolved macroscopic polarization component in the direction of the
probe field, $P(t)$, is computed from the microscopic polarizations
according to semiclassical theory:\cite{book:Haug_Koch}
\begin{eqnarray}\label{Pol_SBE}
    P(t) & = & \frac{1}{w} N_\mathrm{dot} \sum_{n,m} \mu_{nm} \rho_{mn}(t)\, ,
\end{eqnarray}
where $\rho_{nm}$ is the density matrix for localized dot states ($n, m$). Dipole matrix elements are denoted $\mu_{nm}$, $N_\mathrm{dot}$ is the two-dimensional density of the dots in the growth plane, and $w$ is the thickness of the active region.
Within the 3-level, (non-carrier exciting)  approximation one can derive an analytical expression for the susceptibility $\chi (\omega)$ (see \eg Ref.~[\onlinecite{art:hasnain_optical_buffer}]). When both fields are resonant with their respective transitions ($\omega=\omega_p$), the slowdown factor $S$ and the absorption $\alpha$ become:

\begin{eqnarray}
    S(\omega) &=& \left[ \frac{\epsilon_b + \sqrt{\epsilon_b ^2  + \epsilon_{res}}}{2} \right]^{1/2}  \nonumber \\
    &\quad&\times\left[1 + \frac{\hbar \omega}{2 \sqrt{\epsilon_b ^2  + \epsilon_{res}^2}}
      \frac{U_{p} (\Omega_{cc}^2 -
        \gamma_{nr}^2)}{\hbar^2(\gamma_{nr}\gamma_{p} + \Omega_{cc}^2)^2}
    \right] \, . \\
    \alpha(\omega) &=& \frac{\omega}{c_0 n_b } \mathrm{Im}\left[
    \chi(\omega)\right]\nonumber\\
    &=& \frac{\omega}{c_0 n_b } \frac{U_p}{\hbar}\frac{\gamma_{nr} (\Omega^2_{cc} + \gamma_p \gamma_{nr})}
    {(\Omega_{cc}^2 + \gamma_p \gamma_{nr})^2} \, ,
\end{eqnarray}
where $U_{p} = \frac{N_{dot}}{w}|\mu_{p}|^2/\epsilon_0$, $\Omega^2_{cc}= | \mu_{c} |^2 I_c/ 4 \hbar^2 c \epsilon_0 \sqrt{\epsilon_{b}} $, $\epsilon_{res}=  U_{p} / \hbar(\gamma_{p} + \Omega_{cc}^2 / \gamma_{nr})$. $I_c$ is the intensity of the coupling beam, $\mu_{p}$ and $\mu_{c}$ are the dipole moments of the probe and coupling transition, respectively and $\gamma_{p}$ and $\gamma_{nr}$ are the dephasing rates of the polarization components of the probe and the uncoupled non-radiative transition, respectively.
The above expressions displays the significance of maximizing the transition strengths of the two light beams. In case of the slowdown factor the dipole moment of the probe transition determines the largest obtainable slowdown, while the dipole moment of the pump transition is related to the pump power required to reach this maximum. The absorption drops as $1/| \mu_{c} |^2$.

\subsection{Bandstructure}

\begin{figure}[ht]
\begin{center}
\includegraphics[width=7.5cm]{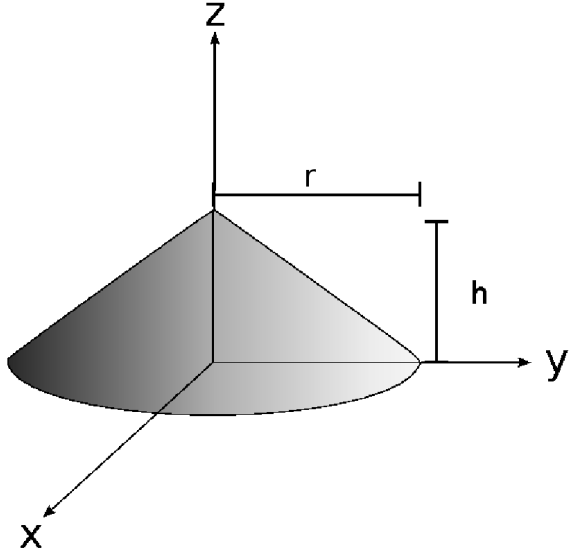}
\end{center}
\caption{The shape of the quantum dots under consideration.}
\label{cone}
\end{figure}

The dipole moments which govern the EIT characteristics are
determined using eight-band $\vec{k}\cdot\vec{p}$ theory including strain effects.
We study the conical
quantum dots shown in Fig.~\ref{cone}. The dot material is InAs and
the barrier material is GaAs.
Both InAs and GaAs are zincblende materials so in order to reduce
the problem to a two dimensional model we disregard anisotropy effects.
This entails that we disregard phenomena such as
piezoelectricity and atomistic anisotropic
effects~\cite{Fonoberov,Barettin}. The atomistic anisotropic effects
have been investigated by Bester and Zunger~\cite{Bester2005}
showing that they lead amongst other things to a splitting of states
which in our model are degenerate. This splitting is however less
pronounced for cylindrical shaped quantum dots as studied here.
Recently it has been shown that second order piezoelectric terms
effectively cancel linear piezoelectric effects for cylindrical
shaped quantum dots. ~\cite{Bester2006,Bester2006PRL,Bimberg1} We
have checked the isotropic assumption and
found a maximum error for the strain fields of $8\%$ along the $z$ axis, going rapidly to
zero at the edges of the dot.

 We investigate both
 the effect of strain and band mixing. The effect of band mixing is
determined by comparing the eight-band results with one-band
model results for the conduction band and the heavy holes.
Due to the lattice mismatch present in the systems under consideration in this
paper (InAs/GaAs quantum dots) the materials will be strained.
In the eight-band model the wavefunction $\psi_n$ of the localized dot state $n$ is given by a linear combination of the eight
 Bloch states weighted by an envelope function,
\begin{align}
 \psi_n= \sum_{i=1}^8 \phi_i^{(n)} u_i     ,
      \end{align}
where $\phi_i^{(n)}$ are the envelope function and $ u_i $ are the
Bloch states. We use the eight-band Hamiltonian described in Ref.~[\onlinecite{Pokatilov}],
while for the strain-dependent part
we follow Ref.~[\onlinecite{Yong}].

In the following we explicitly present the one-band models as due to
their simplicity these are most open to interpretation. The one-band
eigenvalue equation for the envelope functions reads $H_\alpha
\phi_\alpha=E_\alpha \phi_\alpha$ where $H_\alpha$, $\phi_\alpha$
and $E_\alpha$ are the  Hamiltonian, the envelope wave function and
the energy, respectively. $\alpha$ denotes either the conduction
band or the heavy holes, i.e., $\alpha=\{e,hh\}$. The Hamiltonian is
\bea H_\alpha=
H_{\alpha}^{k}(\vec{r_\alpha})+H_\alpha^{b}(\vec{r_\alpha})+H^{\epsilon}_{\alpha}(\vec{r_\alpha})
, \eea where $H_\alpha^{k}(\vec{r}_\alpha)$ is the kinetic part,
$H_\alpha^{b}(\vec{r_\alpha})$
 is the energy of the unstrained band edge, and $H^{\epsilon}_{\alpha}(\vec{r_\alpha}) $
 is the strain dependent part. The solutions are spin
 degenerate in the one-band model, i. e., $\psi_e^{\uparrow} = \phi_e |S
 \uparrow \rangle$ and $\psi_e^{\downarrow} = \phi_e |S  \downarrow \rangle$, and
$\psi_{hh}^{\uparrow} = \phi_{hh} |hh  \uparrow \rangle$ and $\psi_{hh}^{\downarrow} = \phi_{hh} |hh  \downarrow \rangle$.

In the one-band model the strain
Hamiltonian for a zincblende crystal structure is given by
%%%%%%%%%%%%%%%%
\begin{eqnarray}
H^{\epsilon}_{e}(\vec{r_e})    &=&  a_c(\vec{r})\varepsilon_{H}(\vec{r}) ,\\
H^{\epsilon}_{hh}(\vec{r_{hh}})&=& - a_v(\vec{r})\varepsilon_{H}(\vec{r})+ \frac{b}{2}\varepsilon_{B}(\vec{r}),
\end{eqnarray}
%%%%%%%%%%%%%%%%
where $a_c$ ($a_v$) and $b$ are
the conduction (valence)-band hydrostatic deformation potential and $b$ is the shear deformation
potential,~\cite{Bir} while the hydrostatic and biaxial strain components read
%%%%%%%%%%%%%%%%
\begin{eqnarray}
\varepsilon_{H}(\vec{r}) & = &  \varepsilon_{xx}(\vec{r})+ \varepsilon_{yy}(\vec{r})+ \varepsilon_{zz}(\vec{r}), \label{hydrostatic}\\
\varepsilon_{B}(\vec{r}) & = & \varepsilon_{xx}(\vec{r})+
\varepsilon_{yy}(\vec{r})-2 \varepsilon_{zz}(\vec{r}) \label{biaxial},
\end{eqnarray}
%%%%%%%%%%%%%%%%
where $ \varepsilon_{ik}$ is the strain tensor.
The strain
fields are found by minimizing the elastic strain energy.~\cite{Fonoberov}

\subsection{Dipole moments}
The momentum matrix element $\vec{p}_{nm}$ is given by:
\begin{align}
\vec{p}_{nm}
&\equiv \langle \psi_{n} | \vec{p}\,  | \psi_{m}\rangle\notag\\
&=\sum_{i,j=1}^N \bigg( \langle \phi_i^{(n)} | \vec{p}\,  | \phi_j^{(m)}\rangle \delta_{ij}
                               + \langle \phi_i^{(n)} |   \phi_j^{(m)} \rangle \langle u_i | \vec{p} \,| u_j \rangle \bigg)\notag \\
 & \equiv   {\vec{p}_{nm}}^{\,(\phi)} + {\vec{p}_{nm}}^{\,(u)},
\end{align}
where $N=1$ or 8. $ {\vec{p}}^{\,(\phi)} $ and ${\vec{p}}^{\,(u)}$
are the envelope and the Bloch parts of
 the momentum matrix element, respectively.
%For the one-band model $N=1$  while $N=8$ for the eight-band model.
It is usual to consider only  the Bloch part $\vec{p}^{\,(u)}$ since the
 envelope part $ \vec{p}^{\,(\phi)} $ is usually an order of magnitude smaller. \cite{Bimberg}

As shown in Ref.~[\onlinecite{Burt}], the evaluation of the momentum
matrix element $\vec{p}_{nm}$ is meaningful while dipole matrix
$\vec{\mu}_{nm}$ elements are ill-defined in crystals involving
extended Bloch states. Thus, we first calculate $\vec{p}_{nm}$ and
then use the relation between the momentum and the electric dipole
matrix element given by Ref.~[\onlinecite{Merz}]
\begin{align}
\vec{p}_{nm}&=
\frac{im_0}{e} \omega_{nm} \vec{\mu}_{nm},
\label{dipole}
\end{align}
where $ \hbar\omega_{nm}= E_n - E_m$, $m_0$ is the free electron mass and $e$ is the electronic charge.

For material parameters used in calculations we refer to Ref.~[\onlinecite{Vurga}].

\section{Results: bandstructure and dipole moments}
\begin{figure}[ht]\begin{center}
\includegraphics[width=11.5cm]{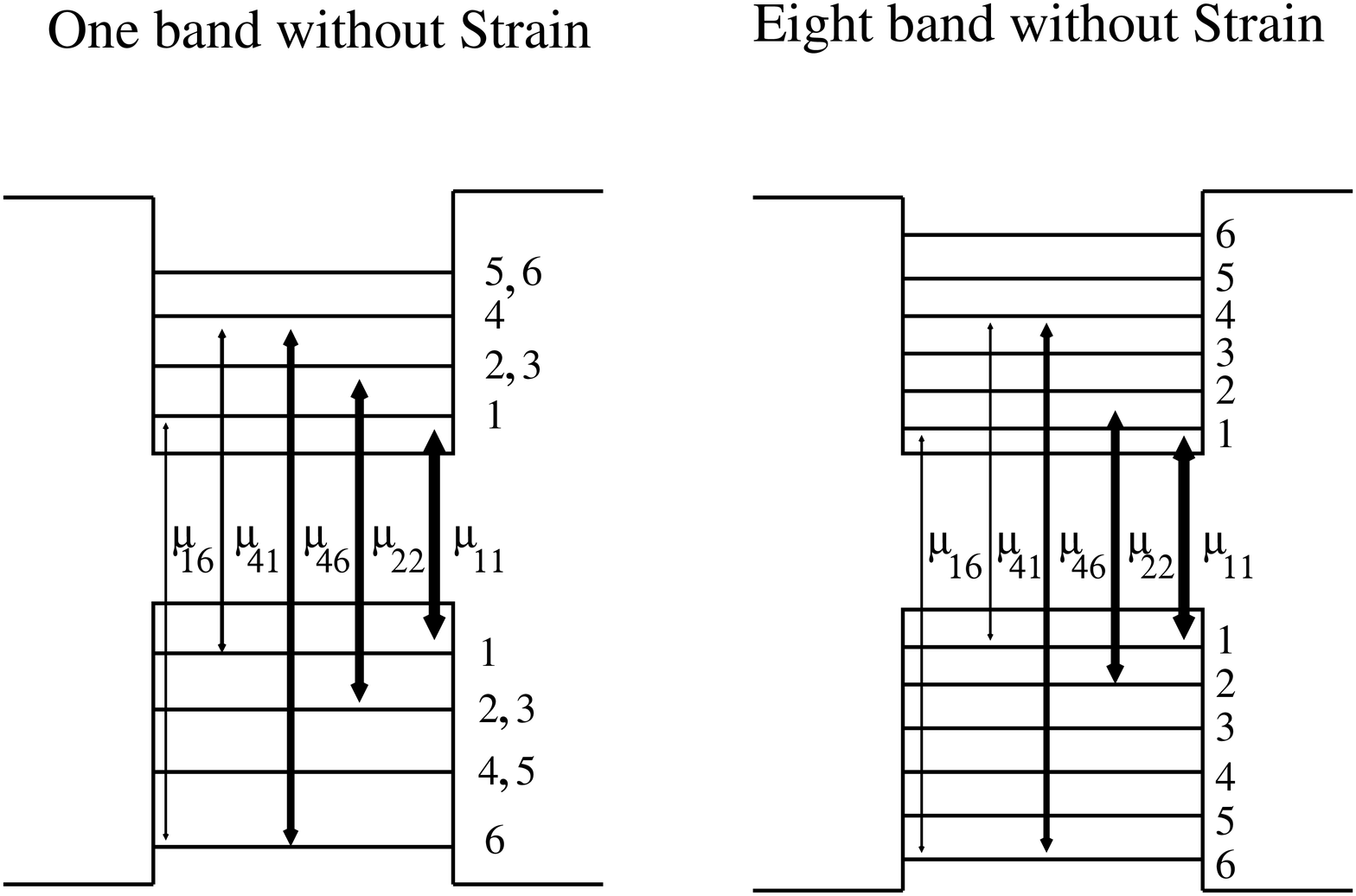}
\includegraphics[width=5.7cm]{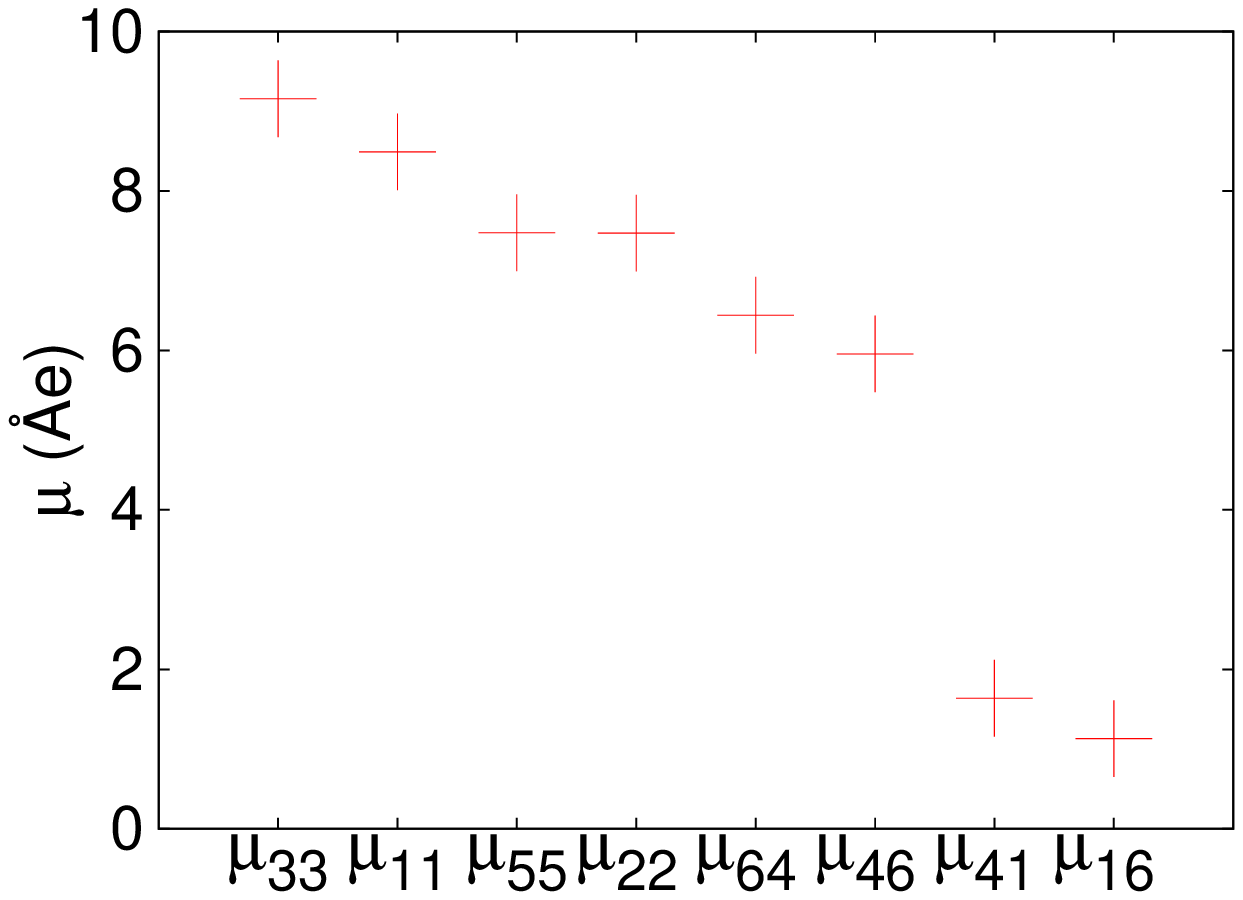}
\includegraphics[width=5.7cm]{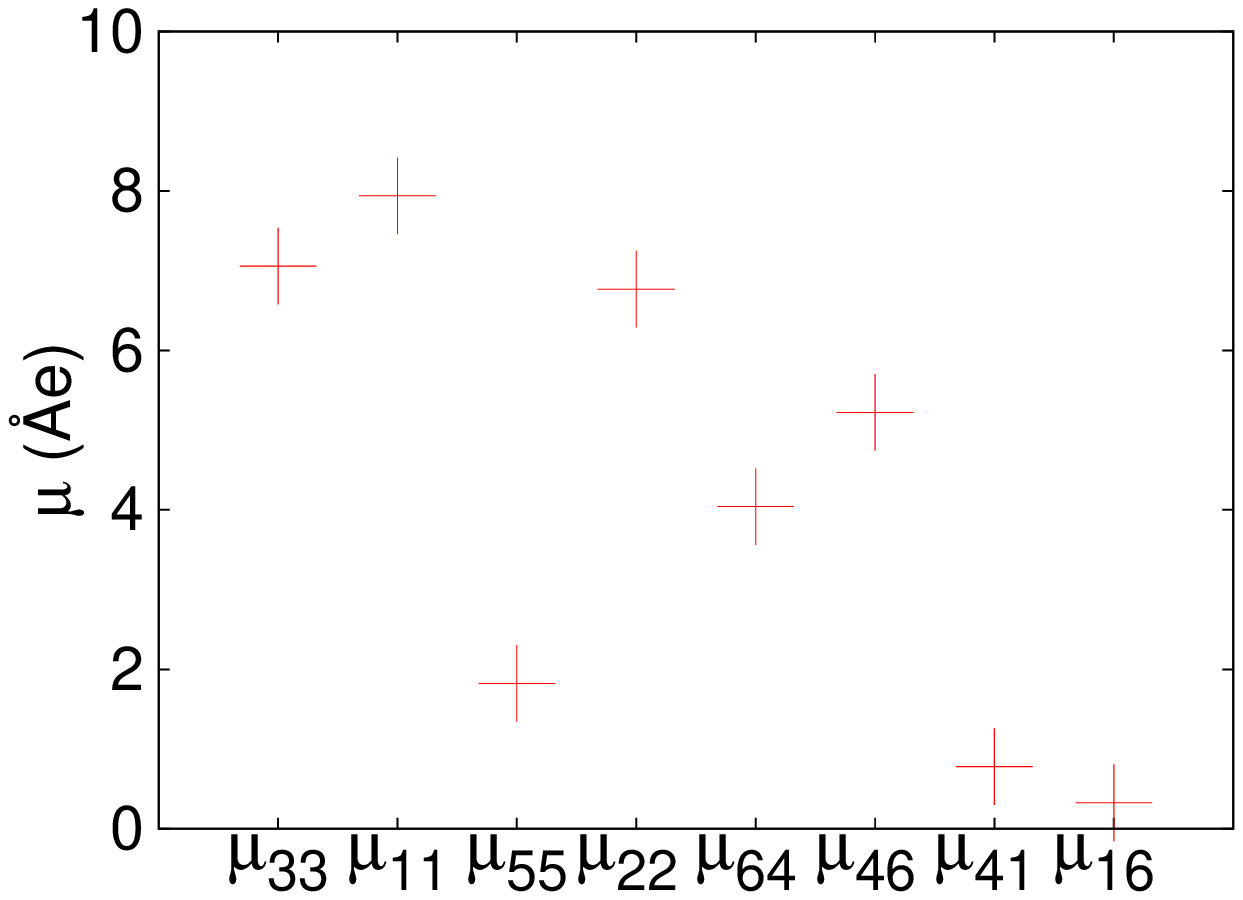}
\end{center}
\caption{Energy levels and the most relevant dipole moments for EIT
(top) and values of the eight strongest interband dipole moments
(bottom) for one-band (left) and eight-band (right) model without
strain for a dot with $h=7.5$ nm.} \label{dipoleset_nostrain}
\end{figure}

\begin{figure}[ht]\begin{center}
\includegraphics[width=11.5cm]{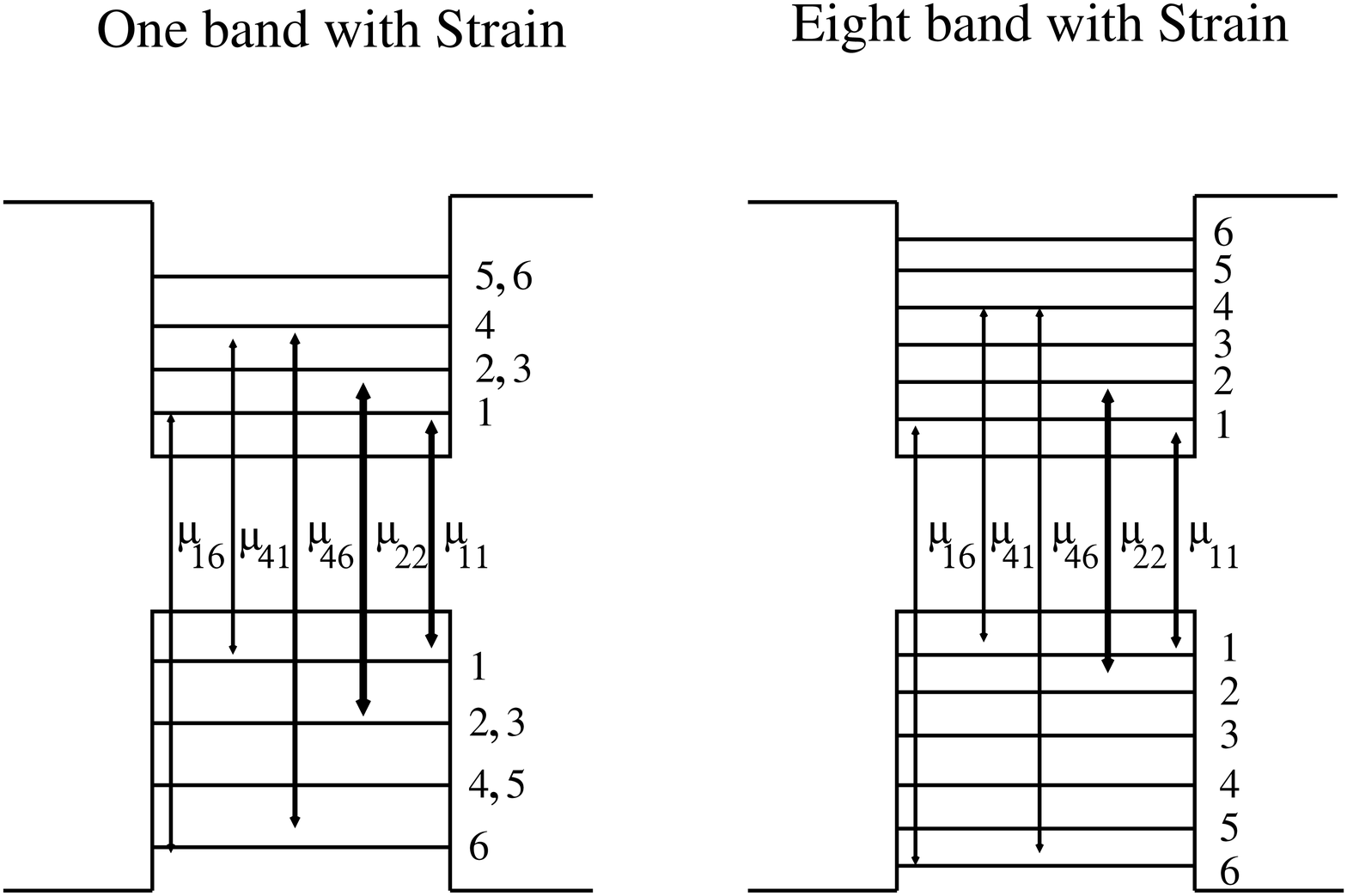}
\includegraphics[width=5.7cm]{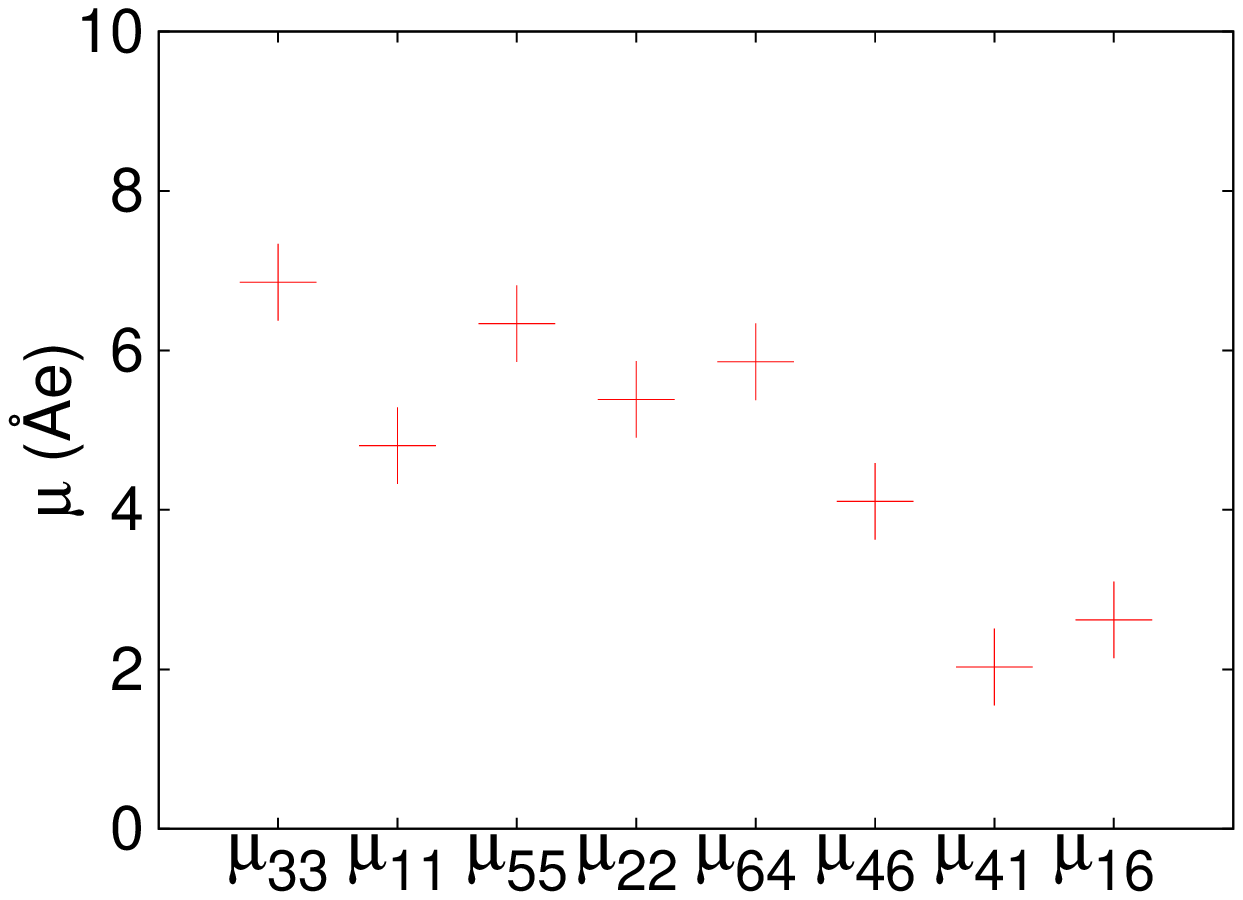}
\includegraphics[width=5.7cm]{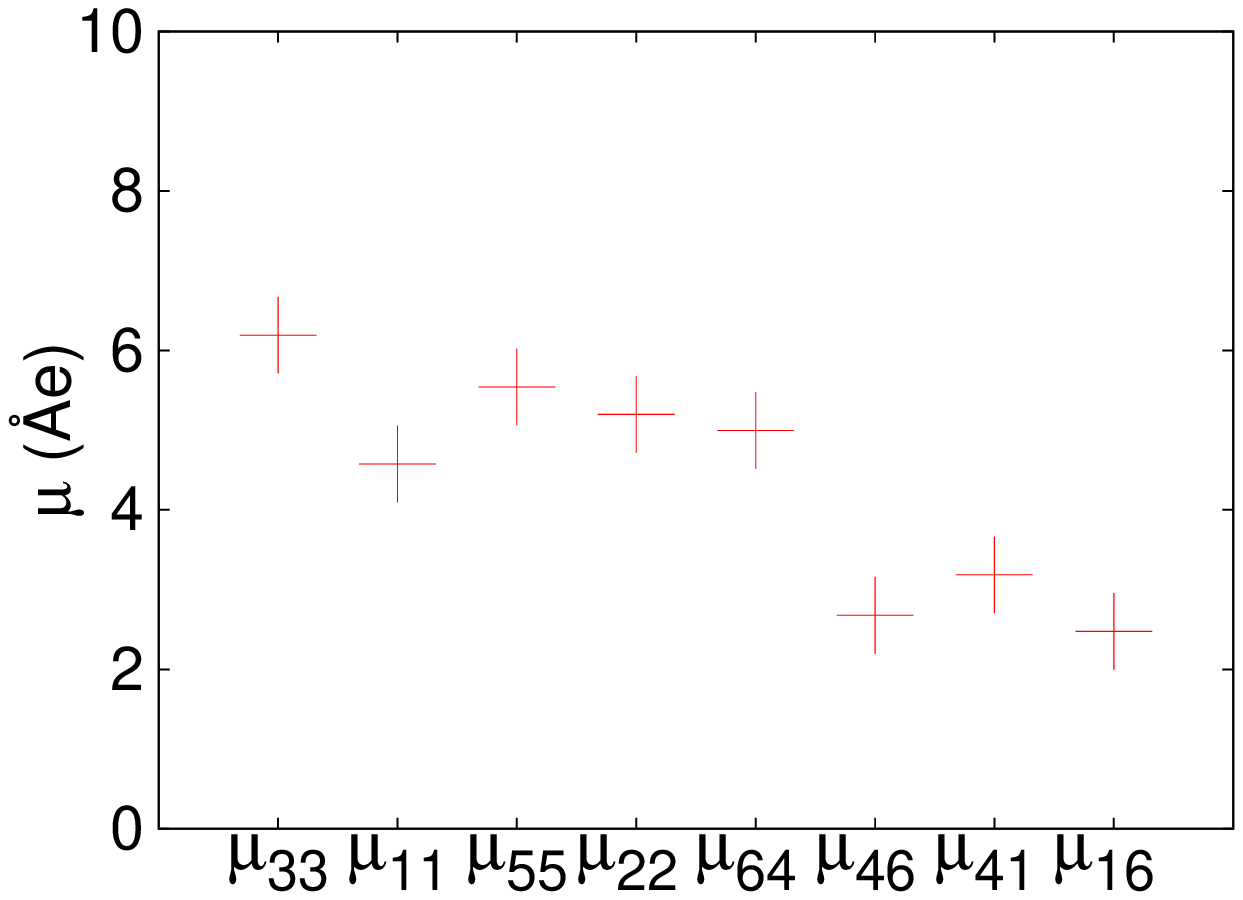}
\end{center}
\caption{Energy levels and the most relevant dipole moments for EIT
(top) and values of the eight strongest interband dipole moments
(bottom) for one-band (left) and eight-band (right) model with
strain for a dot with $h=7.5$ nm.} \label{dipoleset_strain}
\end{figure}

The first results we present are related to a set of conical quantum
dots where the aspect ratio between the radius $r$ and the height
$h$ of the dot has been fixed so that $r=2h$. We focused on the
first twelve bound states for both bands. Since all the states are
at least doubly degenerate (spin degeneracy), we consider six energy
levels (labeled from 1 to 6) for both bands. In the one-band model,
due to the conical quantum-dot symmetry and isotropy assumption
(giving an inversion symmetric model), level 2 and level 3 are
degenerate for both the conduction and the valence band, level 4 and
level 5 are  degenerate for the valence band, and level 5 and level
6 are  degenerate for the conduction band. In the upper part of
Fig.~\ref{dipoleset_nostrain} and Fig.~\ref{dipoleset_strain} we
show the energy levels and the most relevant interband dipole
moments for EIT ($\mu_{22}$ is also included due to its relevance
for other applications) corresponding to a dot with $h=7.5$ nm for
the four different models: one-band model without strain, one-band
model with strain, eight-band model without strain and eight-band
model with strain. Throughout this paper
we consider only right-handed circularly polarized light.


In Figs. \ref{dipoleset_nostrain} and \ref{dipoleset_strain} the thicknesses of the lines indicating the transitions are proportional to the corresponding dipole moments. Obviously, an eight-band model
calculation leads to a higher valence-band density-of-states as
compared to a one-band calculation. Hence, also more interband
transitions result, in a given energy range, when using an
eight-band model. We have chosen to show in the bottom part of
Fig.~\ref{dipoleset_nostrain} and Fig.~\ref{dipoleset_strain} only
the strongest eight dipole-matrix elements for both one-band and
eight-band models. First, we observe that there is a qualitative
agreement between dipole-moment results for the one-band and eight-band models
with strain (Figure~\ref{dipoleset_strain}). This is due to the fact that in the eight-band model
the biaxial strain component of Eq. (\ref{biaxial}) shifts
heavy-holes (light-holes) to higher (lower) energies. As a
consequence the valence-band groundstate and the first excited
states are predominantly heavy-holes like giving rise to a general
better agreement between the one-band model and the eight-band
model. The only notable discrepancies are observed for the EIT
relevant transitions $\mu_{46}$ and $\mu_{41}$. Second, the
inclusion of strain reduces the strength of the dipole moments
significantly. This is because there is a non-trivial influence of
strain on dipole moments. The conduction-band states are only
affected by the hydrostatic strain component of Eq.
(\ref{hydrostatic}) giving rise more or less to a constant shift in
the effective potential inside the dot while the valence band
states, in addition to the hydrostatic strain, are also affected by
the biaxial strain component [see Eq. (\ref{biaxial})]. The latter
component is highly inhomogeneous inside the dot. In the eight-band
model there is a third contribution from $\epsilon_{xz}$ and
$\epsilon_{yz}$ strain components \cite{Yong} but this term is not
as significant as the biaxial of Eq. (\ref{biaxial}).

\begin{figure}[ht,floatfix]
\begin{center}
\includegraphics[width=11.5cm]{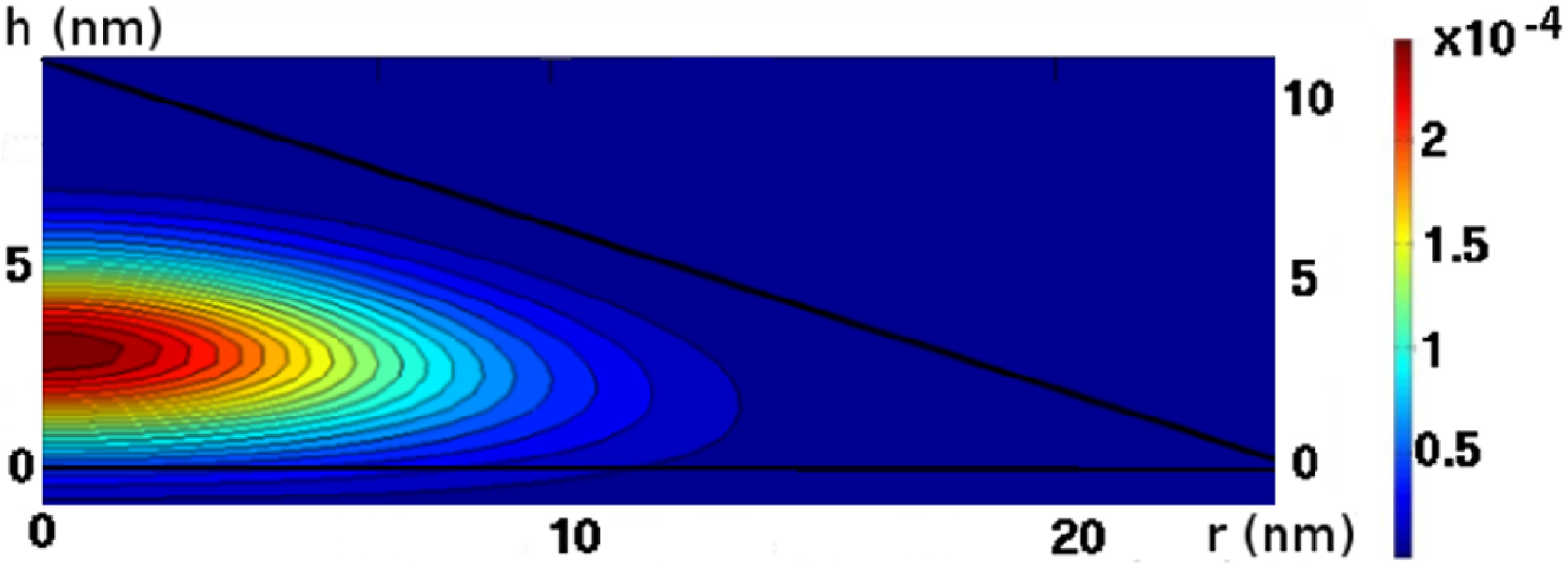}
\includegraphics[width=10.5cm]{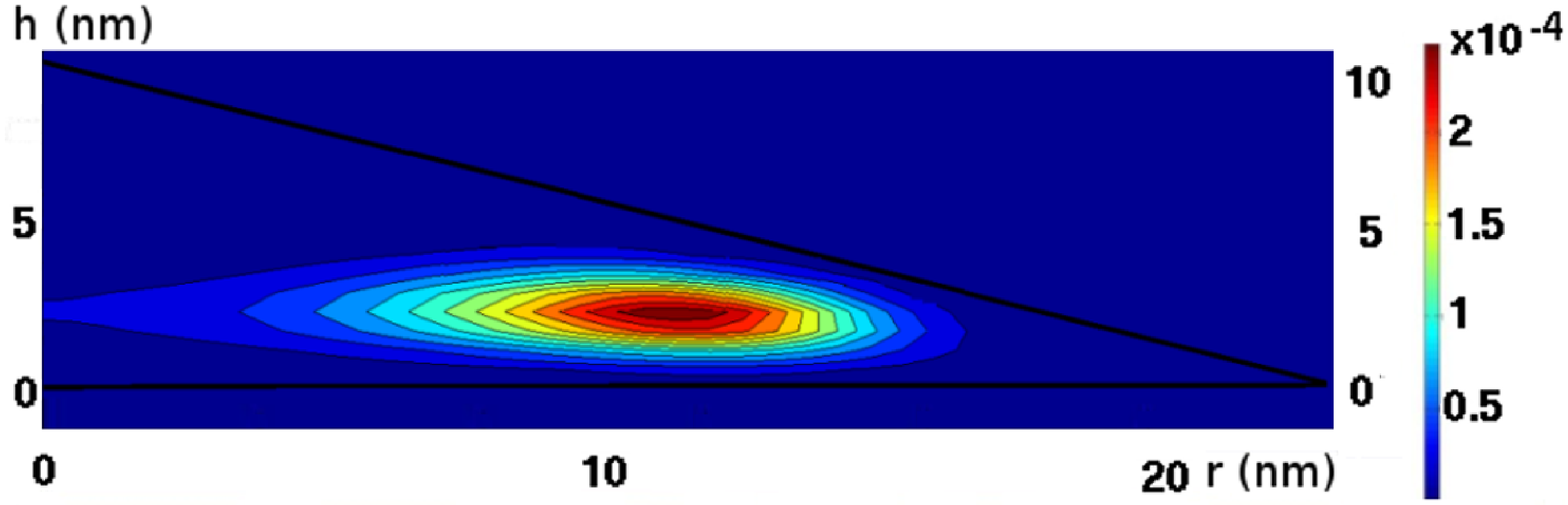}
\end{center}
\caption{Probability density $|\psi|^2$ of the valence-band
groundstate for the eight-band model without (up) and with (down)
strain. The dimensions of the quantum dot are $h=11.5$ nm and $r=23$
nm. } \label{wave}
\end{figure}

In order to
understand the influence of strain on the dipole moment we compare
in Fig.~\ref{wave} the valence-band groundstate probability density
$|\psi|^2$ (eight-band model) with and without the influence of the
strain field for a quantum dot with height $h=11.5$ nm and radius
$r=23$ nm.
While in the
case without strain the groundstate shows a \textit{s-like} shape, the biaxial-strain component
modifies the hole wave function into a toroidal shape moving it away from the center of the
dot where the potential is stronger. This drastically reduces
the overlap between the envelope functions $\phi_i$ of the conduction and
valence band and consequently the corresponding dipole moments.

\begin{figure}[ht]
\begin{center}
\rotatebox{-90}{
\includegraphics [width=7.5cm]{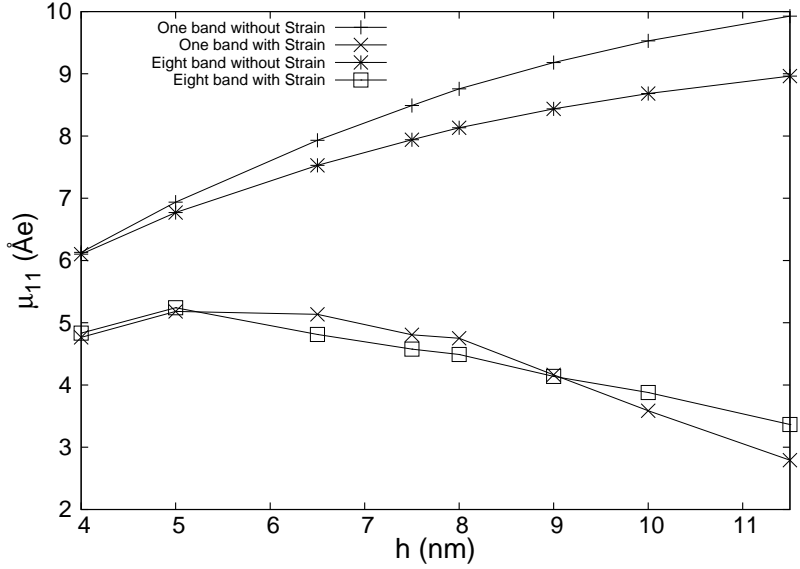}
}
\end{center}
\caption{Interband $\mu_{11}$ dipole moments as a function of $h$
for the four different models.} \label{ss_interband}
\end{figure}

The reduction of the dipole moment due to strain is evident in Fig.~\ref{ss_interband} where we plot the interband
dipole moment $\mu_{11}$ between the conduction- and the valence-band ground states for the four
different models as a function of $h$.
In the models with strain we have a maximum around $h=5$ nm while
the dipole moments for the two models without strain grow
monotonically with increasing height and eventually reach a plateau
value corresponding to the bulk value of $16.8$ \AA e. The
monotonic increase of the dipole moments can be understood based on
Eq. (\ref{dipole}). Without strain the momentum matrix elements
$\vec{p}$ (mainly determined by  ${\vec{p}}^{\,(u)}$) remains
constant with increasing height whereas the energy difference
$\omega_{nm}$ decreases. The presence of strain reduces the overlap
of electron and hole distributions as a result of the increased
displacement of the hole wavefunctions away from the center leading
to the observed decrease in the dipole moments.

\begin{figure}[ht]
\begin{center}
\includegraphics[width=8.5cm]{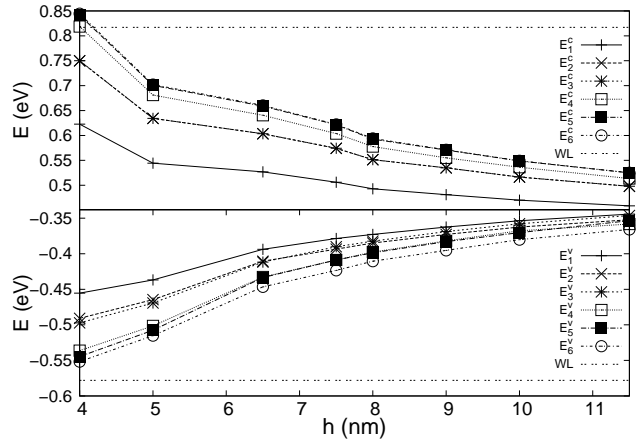}
\end{center}
\caption{Energies of the six considered levels in conduction (top)
and valence (bottom) band as a function of $h$ for eight-band model
with strain. The onset of continuous wetting
layer states (WL) is also indicated (dashed-horizontal line).} \label{energy}
\end{figure}

In Fig.~\ref{energy} we plot the energies of the first six levels in
the conduction (top) and valence band (bottom) as a function of the
height $h$. The wave functions of the confined states are
characterized (in the eight-band model) by eight envelopes weighted
differently depending on the state. As mentioned above, the biaxial strain is
inhomogeneous and this, combined with the differently spatially distributed
envelope functions, leads to a
higher sensitivity against strain as compared to a one-band model.
Further, the inhomogeneity of the strain field grows with volume,
especially for the biaxial-strain component of Eq. (\ref{biaxial}).
These coupled strain-band mixing effects lead to energy crossings in
the valence band. We have also indicated where the wetting layer
continuum starts (WL). The WL dashed lines in Fig.~\ref{energy} are
computed using a Ben-Daniel Duke approach for a 0.5 nm InAs
quantum-well embedded in GaAs \cite{Bastard88}. We observe that only
the last three considered conduction-band levels for the smallest
quantum dot lay above the lower bound of the wetting layer continuum
states.

\begin{figure}[ht]
\begin{center}
\includegraphics[width=8.5cm]{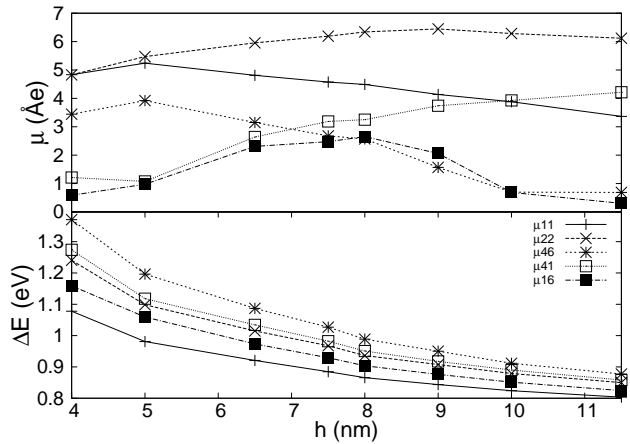}
\end{center}
\caption{Interband dipole moments $\mu$ (top) and relative band
energy difference $\Delta E$ (bottom) as a function of the height
$h$ with fixed aspect ratio $r=2h$.} \label{dipole_volume}
\end{figure}

The most relevant dipole moments for EIT for the eight-band model
are shown in the upper part of Fig.~\ref{dipole_volume} as a
function of $h$. The bottom part gives the related energy
differences $\Delta E$. The strain reduces the dipole-moment
strengths with increasing volume because of the decreasing wave
function overlap (similar to what was found for $\mu_{11}$).

\begin{figure}[ht]
\begin{center}
\includegraphics[width=8.5cm]{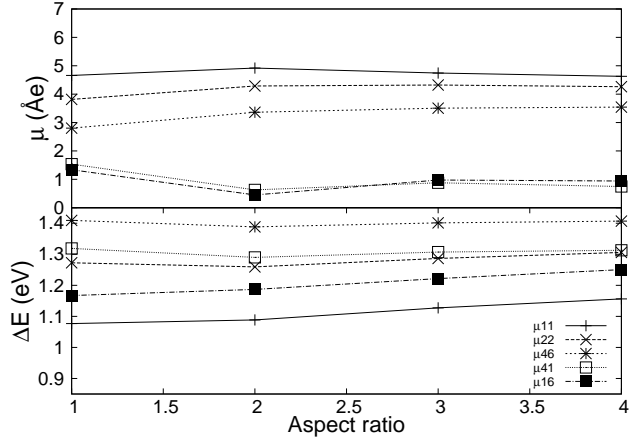}
\end{center}
\caption{Interband dipole moments $\mu$ (top) and relative band
energy difference $\Delta E$ (bottom) as a function of the aspect
ratio at constant volume.} \label{dipole_asp_ratio_variation}
\end{figure}

This geometry effect is indeed mainly a function of the dot volume
as Fig.~\ref{dipole_asp_ratio_variation} shows. Here, we refer to a
second set of quantum dots with different aspect ratios
$(A_{sp}=r/h)$ having the same volume $(V=~226.19~\mbox{ (nm)}^3)$.
The interband dipole moments (top) and relative energy differences
(bottom) are depicted as a function of $A_{sp}$ and evidently
results are rather insensitive to the aspect ratio.

\section{Results: EIT}

\subsection{3 level vs. many levels}

\begin{figure}[ht]
    \begin{center} 
       \includegraphics[width=5.7cm]{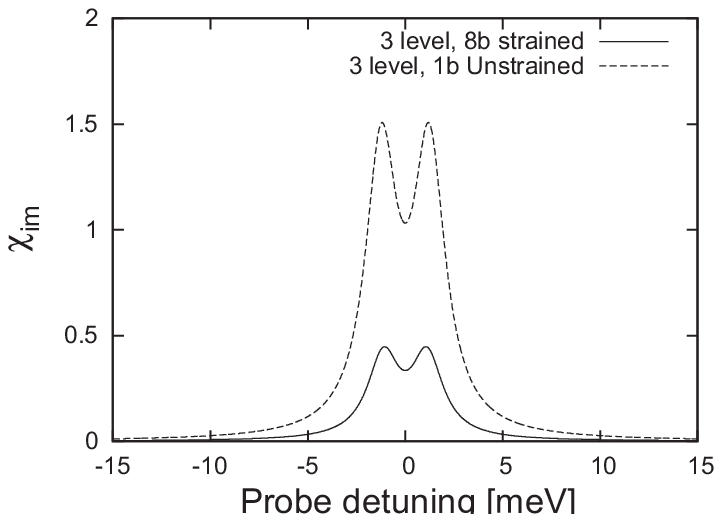}
        \includegraphics[width=5.7cm]{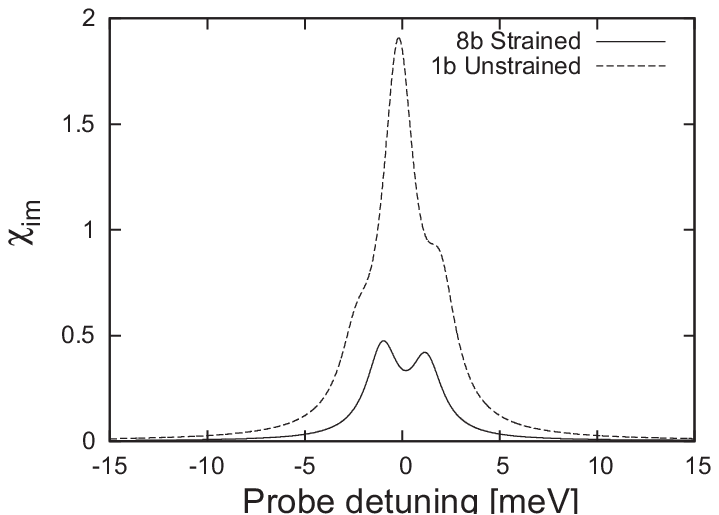}
        \includegraphics[width=5.7cm]{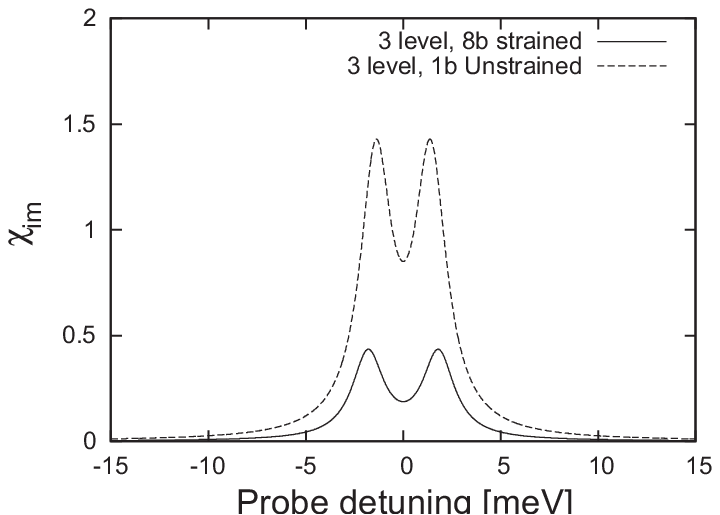}
        \includegraphics[width=5.7cm]{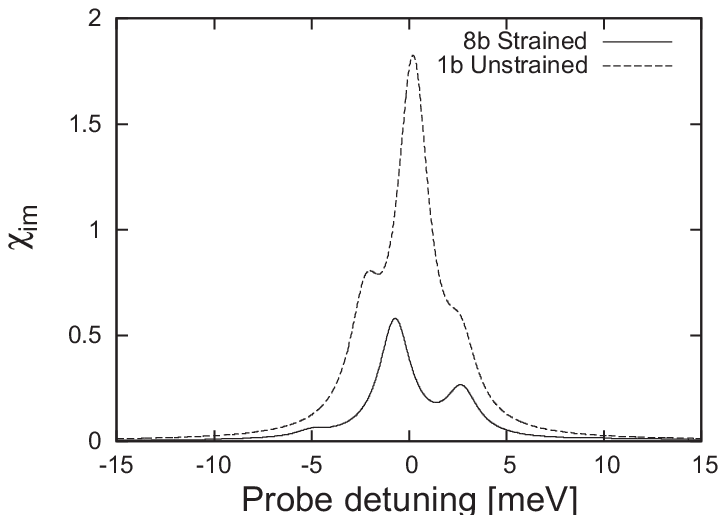}
         % \parbox{0.8\columnwidth}{{\bf Figure 1:}
         \end{center}
\caption{Example of susceptibility calculated for a ladder scheme using the 3 level - or a full multi level approach based on either the eight band strained or one band unstrained model. The results are presented for dots of height 7.5 nm (top panel) and 9 nm (bottom panel), both $A_{sp} = 2$. The reason for the larger features in the one band calculations is that the strength of the probe transition is larger within this model (see Fig.~\ref{ss_interband}).}
        \label{ladder_scheme_issues}
    
\end{figure}

In this section we consider right-handed circularly polarized light for both the coupling and probe fields and investigate the slow light characteristics upon propagation through several stacked layers of QD material. We assume that one can disregard propagation effects in the coupling field. In case of non-carrier exciting schemes, the coupling field is effectively connecting two empty states, thus rendering the transition transparent. In the case where the coupling beam is exciting carriers we consider a propagation through sufficiently thin device so that the coupling intensity remains constant. We use a lattice temperature of 200 K, for which the literature ~\cite{art:Borri_deph_phys_rev_B,art:Borri_dephasing_times} gives dephasing rates around $ 1.5\cdot 10^{12}\; \mathrm{s}^{-1}$. The system response is calculated using a procedure similar to Ref.~[\onlinecite{art:JHN_PRB_EIT}].
First we investigate a ladder-type scheme. The ladder scheme is generally non-carrier exciting. We consider a situation where the probe pulse connects the conduction band/valence band ground state ($\mu_{11}$) and the coupling beam is tuned to the intraband transition in the conduction band between the ground state - and first excited level. This particular setup has been the prime candidate for theoretical scrutiny\cite{art:hasnain_optical_buffer,art:PKN_opt_express} employing a three level approach along with a single band model including only hydrostatic strain. To illustrate the importance of including all energy levels and transitions along with a more detailed bandstructure calculation we show in Fig.~\ref{ladder_scheme_issues} the imaginary part of the susceptibility calculated using a coupling intensity of $1 \; \mathrm{MW}/\mathrm{cm}^2$ for two dot sizes (height 7.5 nm and 9 nm, both $A_{sp} = 2$ ) using either the most simple model (one-band unstrained) or the most complex (eight-band strained).
Most notably the EIT effect, which is recognized as a dip in the absorption spectrum symmetric around the probe frequency and is present in the simple three level models, is absent from three out of four multi-level calculations.
The reason for this is that the additional level structure is dipole coupled to the three level EIT system. In the particular ladder configuration considered here, due to the relatively close spacing between the energy levels, higher lying states are also being addressed by the coupling field thereby adding alternative decay pathways that effectively destroy the destructive interference between the available paths, which is at the origin of EIT. In the one band unstrained calculation the energy levels are almost equidistantly spaced resulting in a peak in stead of a dip in the spectrum. The level structure is modified by strain and the inclusion of additional bandstructure, in fact the multi-level model for the dot of $7.5~\mathrm{nm}$ height indeed displays the EIT effect. In this particular case the higher lying shells are not in as close resonance with the coupling beam and a simple three level treatment is seen to produce similar results as the full multi-level model. The eight-band multi-level calculation for the larger dot ($h=9~\mathrm{nm}$) has reminiscence of the EIT effect but the spectrum is quite far from the "ideal" spectrum predicted by the simple model.
%This alternative pathway is shown in figure (to be added).
In general, ladder schemes in rotationally symmetric dots are, due to the near equidistantly spaced energy levels, likely subjected to the issue of the coupling beam being in resonance with other transitions. Carrier exciting $\Lambda$ or $V$ - schemes connecting only inter-band transitions are due to their larger transition energies much less likely to encounter this problem. Furthermore, the frequency range of inter-band transitions is much easier accessible to industry lasers, than the near infrared required by ladder-type schemes.
\subsection{Size/geometry dependence}
\begin{figure}[ht]
    \begin{center}
    \includegraphics[width=8.5cm]{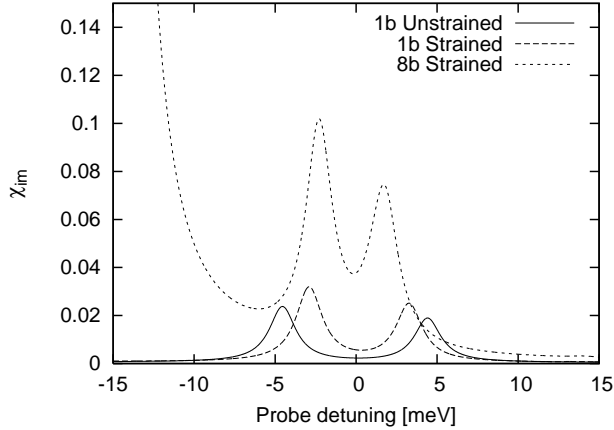}
    \end{center}
    \caption{EIT absorption spectrum calculated using $\mu_{41}$ as probe and $\mu_{46}$ as coupling transition for different models: one-band unstrained (full line), one-band strained (dashed line) and eight-band strained (dotted line). The additional feature in the eight-band strained calculation is due to the nearby lying $\mu_{55}$ resonance.     }\label{1b_vs_8b}  
\end{figure}

%literature mentions several other advantages to using carrier exciting schemes.
The possible interband EIT schemes are illustrated in Fig.~\ref{dipoleset_strain} (top). By combining the shown EIT relevant transitions a variety of different $V$ as well as $\Lambda$-schemes can be accessed. As noted in a previous section there are very few differences between dipole moments predicted by the one-band strained model and the eight-band strained model, however the deviations that do exist, appears for transitions that could be used in connection with EIT. The one-band strained model significantly overestimates $\mu_{46}$, a transition that would otherwise seem attractive to use for EIT. Furthermore $\mu_{41}$ is very small in all but the eight-band strained model. To illustrate the impact of these differences we have in Fig.~\ref{1b_vs_8b} shown absorption spectra for an EIT scheme constructed using $\mu_{41}$ as probe and $\mu_{46}$ as coupling transition.
It is seen that using eight-band strained bandstructure results in an altered EIT behavior. The signature feature, the split peaks, are higher and closer together, a direct consequence of the magnitudes of the involved dipole moments.
These significant differences between the spectra justifies focusing on the results of the more exact eight-band strained bandstructure model. A $V$ and $\Lambda$ scheme can be realized using $\mu_{41}$ as coupling transition and either of the two transitions $\mu_{11}$ or $\mu_{46}$ for the probe field. Two similar schemes can be accessed using the same probe transitions and $\mu_{16}$ for the coupling field.
As is evident from Fig.~\ref{dipole_asp_ratio_variation} the dipole moments are relatively unchanged by varying shape (aspect ratio), we choose to focus on $A_{sp}=2$  since the common probe transitions ($\mu_{11}$ and $\mu_{46}$) are maximized for this geometry and consider volume dependencies.

We see that for small dots ($h<5~\mathrm{nm}$) the dipole moments addressed by the coupling beam are relatively small (on the order of $1~\mathrm{nm}$) and therefore dots of this size are not suitable for EIT operation. As the dot increases in size both coupling transitions become stronger, the two probe transitions, however begins to decrease. The dipole $\mu_{46}$ is always the smaller of the two probe transitions, and the fact that it drops off faster suggests that one should aim at using $\mu_{11}$, ground state to ground state, as probe transition.

\begin{figure}[ht]
    \begin{center}
    \includegraphics[width=8.5cm]{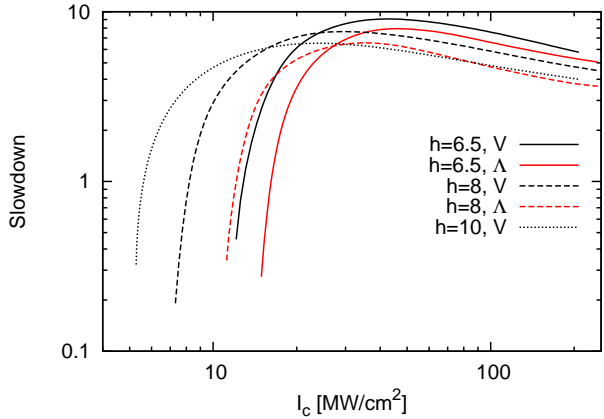}
    \end{center}
    \caption{(Color online) Slowdown versus coupling beam intensity calculated for three different dot sizes. The black curve refers to the $V$ scheme whereas the red curve is for $\Lambda$ configuration. }\label{slowdown}
     \end{figure}

To quantify the above discussion we show in Fig.~\ref{slowdown} the slowdown factor as a function of coupling intensity calculated for three different dot volumes ($h=6.5~\mathrm{nm}, 8~\mathrm{nm}, 10~\mathrm{nm}$) for each of the two remaining EIT schemes (for the $h=10~\mathrm{nm}$ dot, in the case of the $\Lambda$ scheme no EIT behavior is found due to its weak coupling transition. Slowdown results for this dot/EIT setup has therefore been omitted in Fig.~\ref{slowdown}.). For every dot type we see that the $V$-scheme prevails, it has the lowest coupling power requirements and the largest obtainable slowdown value. The first observation can be understood from Fig.~\ref{dipole_volume} where the $V$-scheme coupling transition $\mu_{41}$ is always the larger. Since both schemes utilize the same probe transition one could be inclined to think that they should display the same maximum slowdown value. The reason they do not, is that the coupling beam, although detuned, is exciting carriers into the levels connected by the probe beam. The presence of the carriers serve to block the probe transition. As indicated in the lower part of Fig.~\ref{dipole_volume}, the coupling beam in the $\Lambda$ case is less detuned from the probe transition than in the V case, and thus more carriers are excited into the electron and hole ground states, resulting in a smaller effective probe transition strength.
It is apparent that the best tradeoff between maximum slowdown and coupling intensity is found using the $V$ scheme for dots having a relatively large height, in the region around $8~\mathrm{nm}$. In the literature the $V$ scheme has also been deemed preferable due to its ability to overcome inhomogeneous broadening,\cite{art:PLH_APL_2009} and a favorable carrier redistribution mechanism.\cite{art:JHN_PRB_EIT}

\section{Conclusion}

For the investigated system we have shown that the strain field generally
reduces optical transition strengths as a function of the volume of the
quantum dot for a fixed aspect ratio. This is due to a decreasing overlap of the
involved wavefunctions. This geometry effect has been shown to be a volume
effect rather than a shape effect. Moreover, the combined influence of band
mixing and strain entails state crossings in the valence band, and a
separation of the heavy and light holes. The latter leads to a qualitative
agreement between the lower conduction and upper most valence-band states
computed using the one-band model and the eight-band model. The observed discrepancies for $\mu_{46}$ and $\mu_{41}$ have an impact on the choice of EIT scheme.
We have investigated the importance of including the full energy level structure and optical transitions in an EIT simulation. In particular we find for a ladder type scheme that the additional transitions along with an almost equidistant energy level spacing can severely impair the EIT effect. We have studied the effect of varying dot size and geometry on EIT and identified a $V$-type configuration in a dot of aspect ratio $r=2h$ with height $h \approx 8~\mathrm{nm}$ to have the best possibilities for efficient EIT operation.

\begin{acknowledgements}
This work is supported by the Danish Research Council for Technology
and Production (QUEST). APJ is grateful to the FiDiPro program of
the Finnish Academy.
%and the European Commission through the IST
%project "QPhoton" (Contract No.~IST-29283).
\end{acknowledgements}
%\section*{References}

%\bibliographystyle{iopart-num}
%\bibliography{quantum-coherence-references}

\end{document}